\documentclass[floats,aps,showpacs,superscriptaddress,amsmath,amsymb,prl,twocolumn]{revtex4}
\usepackage{graphicx}
\usepackage{color}
\usepackage{amsmath}
\usepackage{rotating}
\usepackage{dcolumn}
\usepackage{bm}

\begin{document}
\newcommand{\g}{\bf}

\title{Anomalous Hall effect in a two dimensional electron gas with
  magnetic impurities}

\author{Tamara S.\ Nunner}
\affiliation{Institut f\"ur Theoretische Physik, Freie Universit\"at
  Berlin, Arnimallee 14, 14195 Berlin, Germany}
\author{Gergely Zar\'{a}nd}
\affiliation{Institute of Physics, Technical University Budapest,
  Budapest, H-1521, Hungary}
\author{Felix von Oppen}
\affiliation{Institut f\"ur Theoretische Physik, Freie Universit\"at
  Berlin, Arnimallee 14, 14195 Berlin, Germany}

\date{\today}

\begin{abstract}
Magnetic impurities play an important role in many spintronics-related materials. Motivated by this fact, we study the anomalous Hall effect in the presence of magnetic impurities, focusing on two-dimensional electron systems with Rashba spin-orbit coupling. We find a highly nonlinear dependence on the impurity polarization, including possible sign changes. At small impurity magnetizations, this is a consequence of the remarkable result that the linear term is independent of the spin-orbit coupling strength. Near saturation of the impurity spins, the anomalous Hall conductivity can be resonantly enhanced, due to interference between potential and magnetic scattering. 

\end{abstract}

 \maketitle

{\it Introduction.}---In ferromagnetic materials,  the Hall resistance acquires an anomalous
contribution which is proportional to the magnetization of the
sample~\cite{KarplusLuttinger,Smit,Berger}. 
Although this anomalous Hall effect (AHE) has become a standard tool to
determine the magnetization of ferromagnets and has been known for more than a
century, its mechanism is still under debate. The interest in the
origin of the AHE has recently been renewed due to its close relation
to the spin Hall effect.
Particular attention has been paid to intrinsic mechanisms~\cite{KarplusLuttinger}, where the
spin-orbit interaction modifies the
band-structure. This is in contrast to
extrinsic mechanisms~\cite{Smit,Berger,WoelfleMuttalib}, where the spin-orbit
interaction appears only in
the impurity potential. Surprising results have been found even
for  simple systems such as a two-dimensional electron gas (2DEG)
with Rashba spin-orbit interaction
where it turns out that in the presence of pointlike potential
disorder the AHE vanishes when both of the spin-split
bands are occupied~\cite{Culcer,Dugaev,Sinitsyn05,Onoda,Inoue,Borunda,AHElong}.

So far, most treatments of the intrinsic AHE have only considered the
presence of potential impurities. However, in many materials, which
are of interest for spintronics applications such as diluted magnetic
semiconductors, magnetic impurities play a fundamental role. Although
extrinsic mechanisms for the AHE based  on scattering by magnetic
impurities have been suggested~\cite{Fert},
little is known about the influence of magnetic impurities in
materials with strong spin-orbit interaction, where the AHE is induced
by intrinsic mechanisms.
To learn about the role of magnetic impurities, we avoid the complexities of realistic band structures
and instead provide a thorough analysis of 
a simple model: a 2DEG with Rashba spin-orbit interaction where the presence of magnetic impurities is known to induce a finite AHE even when both spin-split bands are occupied \cite{Inoue}.
We find that surprisingly rich physics emerges even in this model system. 

We expect that our model can be tested experimentally in magnetically doped 2DEGs~\cite{Smorchkova},
once they are produced with asymmetric confinement potentials. This may be realized in heterojunctions, 
made of a material with large spin-orbit coupling. At present, a robust AHE has been observed in a magnetically doped 2DEG with {\em weak} spin-orbit coupling, based on a modulation-doped quantum well of Zn$_{1-x-y}$Cd$_y$Mn$_x$Se 
($x \sim 0.02, y \sim 0.12$) sandwiched between ZnSe barriers \cite{Cumings}.
  
To construct a theory for the AHE in magnetically doped 2DEGs 
with Rashba spin-orbit coupling, we use the Hamiltonian
\begin{equation}
H=\frac{{\bf p}^2}{2m} + \alpha (p_y \sigma_x - p_x \sigma_y)
 +\sum_i [V + J {\bf S}_i \cdot {\bm \sigma}] \delta({\bf r}-{\bf R}_i) \,.
\label{eq:Hamiltonian}
\end{equation}
Here, $\alpha$ denotes the strength of the spin-orbit interaction. The impurities at positions
${\bf R}_i$ affect the conduction electrons through a potential $V\delta({\bf r}-{\bf R}_i)$ and 
an exchange coupling $J {\bf S}_i\cdot{\bm \sigma}$. (Here, ${\bf S}_i$ denotes the impurity and ${\bm \sigma}$
the electron spin.) Our analysis below includes interference of 
amplitudes from potential and magnetic scattering, and 
accounts for all significant contributions to the AHE, including skew scattering. 
Remarkably, we find that the AHE deviates significantly from the conventional linear
magnetization dependence except for the region of very
small polarization of the impurity spins. For strong impurity polarization,
we obtain a resonant behavior of the anomalous Hall conductivity. This enhancement arises
from an interference-induced suppression of the scattering rate for the minority carriers.

{\it Anomalous Hall conductivity.}---Our theory is based on the Streda-Kubo approach which decomposes the Hall conductivity 
$\sigma_{yx}=\sigma_{yx}^{I}+\sigma_{yx}^{II}$ into a Fermi-surface contribution 
$\sigma_{yx}^{I}$ and a contribution $\sigma_{yx}^{II}$ from the entire Fermi sea~\cite{Streda}.
The AHE is dominated by the Fermi surface contribution
\begin{equation}
\sigma_{yx}^{I}\!=\! \frac{e^2}{4 \pi V} 
{\rm Tr} \! \left [ 
v_y G^R_{\epsilon_F} v_x (G^R_{\epsilon_F}\!\!-\!G^A_{\epsilon_F})
\!-\! v_y (G^R_{\epsilon_F}\!\!-\!G^A_{\epsilon_F}) v_x G^A_{\epsilon_F}
  \right ],
\end{equation}
where $G^{R,A}$ denotes the advanced and retarded Green functions and 
the velocity operators are given by $v_{x,y}=p_{x,y}/m \mp \alpha \sigma_{y,x}$.
It can be shown that the contributions of $G^R G^R$ and $G^A G^A$ are of higher order 
in the disorder scattering rate $\Gamma$ \cite{Sinitsyn}, leaving only the mixed $G^RG^A$ terms to be 
considered. 
Regarding $\sigma_{yx}^{II}$ it is known that it
vanishes for $\epsilon_F > h$ in clean systems~\cite{AHElong}, where
$h$ is the effective magnetic field.  
This remains true also in the presence of magnetic impurities as long as vertex
corrections are neglected, which contribute only
to higher order in $\Gamma$~\cite{Sinitsyn}.

To compute the anomalous Hall conductivity, we treat the exchange
interaction at a mean-field level, average over the impurity 
potential, and  incorporate disorder-effects perturbatively.
Within the mean field approximation, the
exchange interaction leads to an additional internal Zeeman field
acting on the electrons, which is typically larger than the applied
magnetic field $h_{\rm ext}$. Thus, the electrons are subject to an 
average effective Zeeman field $h \simeq \epsilon_J \langle S_z \rangle/S$, where $\epsilon_J = n_i JS$
is the maximal exchange field and $n_i$ denotes the concentration of impurities.
At mean-field level, the nonzero impurity-spin expectation values are given by 
$\langle S_z \rangle=-\partial\ln Z /\partial(\beta \tilde h )$, 
$\langle S_z^2 \rangle=(1/Z) \partial^2 Z/\partial (\beta \tilde h)^2$
in terms of the Brillouin partition function  
$Z=\sinh (\beta J \tilde h  (2S+1)/2)/\sinh (\beta \tilde h /2)$ and 
the effective field $\tilde h= n J \langle \sigma\rangle$ acting on the impurity 
spins. Here, $n$ and $\langle \sigma\rangle$ denote the density and polarization
of the electrons. We shall not determine $\langle S_z \rangle$
self-consistently. Instead, we treat it as a {\em
  parameter} which also determines the impurity spin fluctuation $\langle S^2_z
\rangle$ within this mean field approach.

Let us now consider the effects of disorder. In the mean field approximation,
the dispersion of the two subbands is given by
$E_{k,\pm}= k^2/2m \pm \lambda_k$ with $\lambda_k=\sqrt{h^2 +\alpha^2 k^2}$ and the retarded
Green function takes the form $G^{0,R} = G^{0,R}_0 + \sum_{i=x,y,z} G^{0,R}_i \sigma_i$ with
\begin{eqnarray}
\label{eq:GreensFunction}
&& \!\!\!\!\!G^{0,R}_0 \!=\! \frac{1}{2} ( G^{0,R}_+ \!\!+ G^{0,R}_- )\,\, ,\,\, 
G^{0,R}_z \!=\! - \frac{ h}{2 \lambda_p} (G^{0,R}_+ \!\!- G^{0,R}_-)
\\
&&\!\!\!\!\!G^{0,R}_{x/y} \!=\! \pm \frac{\alpha p_{y/x}}{2  \lambda_p}   
(G^{0,R}_+ \!\! -G^{0,R}_-)  \,\, ,\,\, 
G^{0,R}_\pm \!=\! \frac{1}{\omega-E_{p\pm}+i0^+} \nonumber \,.
\end{eqnarray}
Deriving the retarded self-energy of the 2DEG within the Born approximation, we obtain
$\Sigma^R = -i (\Gamma + \Gamma_z\sigma_z)$ with
\begin{eqnarray}
&&\Sigma^R = - \frac{i n_i}{4} \left\{ [V^2 +J^2 S(S+1)+ 2V J \langle S_z \rangle \sigma_z ]
K_1 \right. \\
&& \left. -  
\left[ 2V J \langle S_z \rangle + [V^2 + J^2 (2\langle S_z^2 \rangle - S(S+1))] \sigma_z \right]
h K_2 \right\}\, . \nonumber
\label{eq:SelfEnergy}
\end{eqnarray}
Here, $K_1=(\nu_+ + \nu_-)$ and $K_2={\nu_+}/{\lambda_+} -
{\nu_-}/{\lambda_-}$ where 
$\nu_\pm=m \lambda_\pm/ \sqrt{h^2+2 \epsilon_{\rm so} \epsilon_F +
(\epsilon_{\rm so})^2}$, where $\epsilon_{\rm so} = \alpha^2 m$
is related to the density of states
of the two bands at the Fermi energy $\epsilon_F$ and $\lambda_{\pm}=\lambda_{k_\pm}$ 
with $k_\pm$ being the corresponding Fermi wavevectors. 
If only pointlike potential scatterers are present and scattering from magnetic impurities is neglected, the anomalous
Hall conductivity vanishes except for the extreme limit $\epsilon_F < h$. 
In this paper, we focus on the more realistic regime of $\epsilon_F > h$ where {\it both} subbands
are partially occupied and where it is the magnetic impurities which induce a
nonzero anomalous Hall effect.

When both subbands are occupied, one finds $K_2=0$ and hence a significant simplification
of the self energy in Eq.\ (\ref{eq:SelfEnergy}). The impurity averaged Green function also takes the form
of Eq.~(\ref{eq:GreensFunction}), with the replacements $E \to E+i\Gamma$ and $h\to h+i\Gamma_z$. In the
following, we assume weak disorder scattering in the sense $\Gamma_z \ll \lambda_p$ such that 
\begin{equation}
G^R_\pm\simeq \frac{1}{\omega-E_{p\pm}+i\Gamma_{\pm}} \,, \quad
\Gamma_\pm = \Gamma \mp \Gamma_z \frac{h}{\lambda_\pm}.
\label{eq:Gammapm}
\end{equation}

\begin{figure}[t]
\begin{minipage}{.55\columnwidth}
\includegraphics[height=1.05cm]{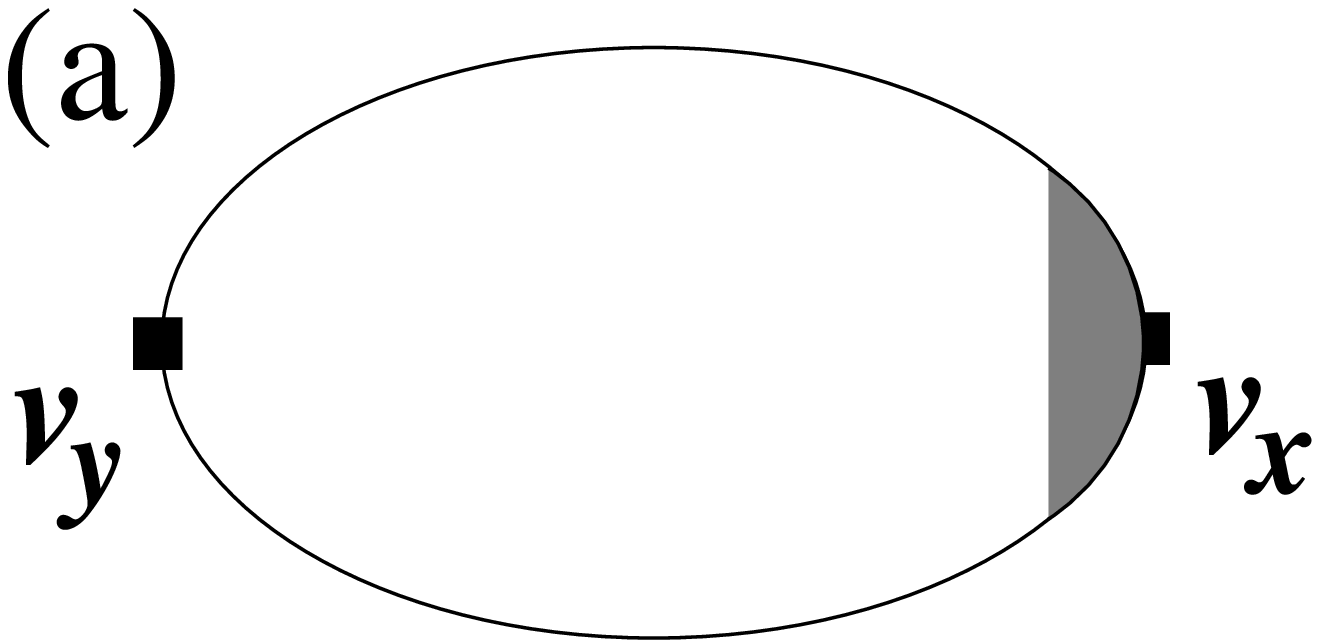}\\
\includegraphics[height=1.1cm]{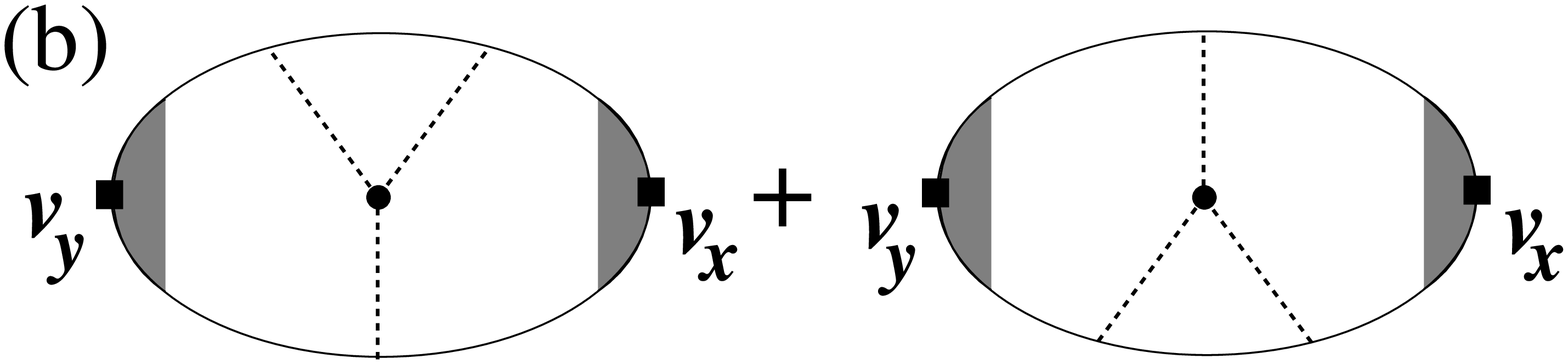}
\end{minipage}
\begin{minipage}{.4\columnwidth}
\includegraphics[height=.9cm]{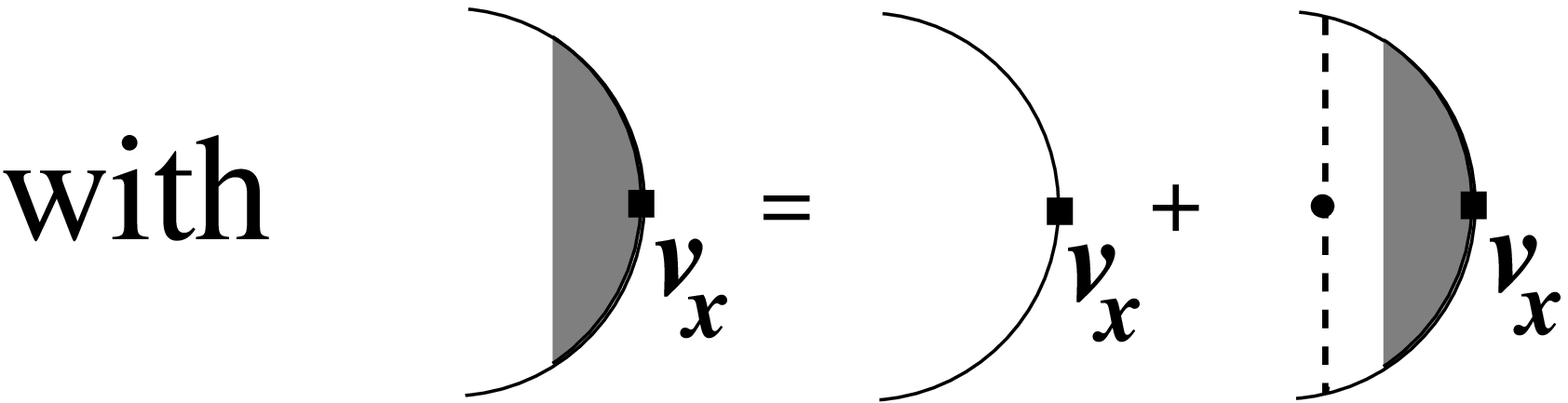}
\end{minipage}
\caption{Anomalous Hall conductivity: (a) Ladder contribution and (b)
  skew scattering contribution, in analogy to
  Refs.~\cite{Borunda,AHElong} we only consider diagrams with a single
  third order impurity vertex with both external current vertices 
  renormalized by ladder type vertex corrections.}   
\label{fig:SigmaDiag}
\end{figure}

The anomalous Hall conductivity $\sigma_{yx}^I$ can now be computed as the sum of
the ladder diagrams $\sigma_{yx}^{I,l}$ (see Fig.~\ref{fig:SigmaDiag}(a)) and the skew-scattering contribution 
$\sigma_{yx}^{I,s}$ (see Fig.~\ref{fig:SigmaDiag}(b)). 
Using $n_i \tilde V^2= n_i (V^2 -J^2 \langle S_z^2 \rangle )$, a lengthy but standard evaluation of the diagrams yields
\begin{equation}
\sigma_{yx}^I= \sigma_{yx}^{I,l}
             + \sigma_{yx}^{I,s}
\end{equation}
with 
\begin{eqnarray}
\label{eq:sigmaAHE}
\sigma_{yx}^{I,l} \!\!\!&=&\!\!\! \frac{e^2}{2\pi} 2 \alpha 
(\gamma_1 + i I_3) \\
\!\!\!&-&\!\!\! \frac{e^2}{\pi}
\frac{n_i \tilde V^2
\!\left( 2 \gamma_1 \gamma_2 (1\!\!-\!\!n_i \tilde V^2 I_1)
\!\!+\! i n_i \tilde V^2 I_2 (\gamma_2^2 \!\!-\!\! \gamma_1^2) \right)}
{(1 -  n_i \tilde V^2 I_1)^2-(n_i \tilde V^2 I_2)^2} 
\nonumber \\
\sigma_{yx}^{I,s} \!\!\!&=&\!\!\! \frac{e^2}{2\pi} 
\frac{2 n_i m J \langle S_z \rangle (V^2 -J^2S(S+1))  \gamma_2^2}
{(1-n_i \tilde V^2 I_1)^2} 
\nonumber
\end{eqnarray}
where 
\begin{equation}
\gamma_1 = i (I_3 -\alpha I_2) \, \quad
\gamma_2 = I_4 -\alpha I_1
\end{equation}
and
\begin{eqnarray}
&I_1& \approx \frac{1}{8} \left (
          \left(1-\frac{h^2}{\lambda_+^2} \right) \frac{\nu_+}{\Gamma_+}
          +\left(1-\frac{h^2}{\lambda_-^2} \right) \frac{\nu_-}{\Gamma_-} 
\right ) \\
&I_2& \approx - \frac{i}{4} \left ( 
\frac{\nu_+ h}{\lambda_+^2} + \frac{\nu_- h}{\lambda_-^2}
-\frac{\Gamma_z}{\Gamma_+} \frac{\nu_+ \alpha^2 k_+^2}{\lambda_+^3}  
+\frac{\Gamma_z}{\Gamma_-} \frac{\nu_- \alpha^2 k_-^2}{\lambda_-^3}  
\right ) \nonumber \\
&I_3& \approx -\frac{i}{4} \alpha \Gamma_z \left (
    \frac{\nu_+}{\Gamma_+ \lambda_+} \left (
    \frac{\epsilon_F}{\lambda_+} -1 \right)
   + \frac{\nu_-}{\Gamma_- \lambda_-} \left (
    \frac{\epsilon_F}{\lambda_-} +1 \right)
\right ) \nonumber \\
&I_4& \approx - \frac{1}{4} \alpha \left (
\epsilon_F \left( \frac{\nu_+}{\Gamma_+ \lambda_+}
                 -\frac{\nu_-}{\Gamma_- \lambda_-} \right )
-\left( \frac{\nu_+}{\Gamma_+}+\frac{\nu_-}{\Gamma_-} \right) 
\right ) \, . \nonumber
\label{eq:integrals}
\end{eqnarray}
Here we have taken the weak scattering limit of $\sigma_{yx}^{I,s}$.
As a result of several cancellations these expressions 
turn out to be formally similar to the ones of Ref.~\cite{AHElong},
derived for purely potential scatterers. The main
differences are that additional terms proportional
to $\langle S_z \rangle$ appear in the self-energy and that $n_i V^2$ is
replaced by $n_i \tilde V^2$ in the vertex corrections. 
Furthermore,  $V^2$ is replaced by
$V^2-J^2S(S+1)$ in the numerator of the skew
scattering contribution. 

We remark that the expression
for the ladder contribution reduces to the
results of Inoue {\it et al.}~\cite{Inoue}, i.e., 
$\sigma_{yx}^{I,l} \approx e^2 \epsilon_{\rm so} \epsilon_F^2 \delta^2/(2 \pi h^3)$
with $\delta=\Gamma_z/\Gamma \approx 2J \langle S_z \rangle/V$,
when taking the special limit of small spin-orbit interaction, large Fermi energy
$\Delta_{\rm so} \ll h \ll \epsilon_F$ with $\Delta_{\rm so}=\alpha k_F$, as well as small
exchange component of the impurity potential. 
Here, however, we focus on the remarkably rich behavior of the anomalous
Hall conductivity beyond this special limit. Although our treatment is
set up for quantum spins, we shall restrict most formulas to classical
spins in the remainder of the paper. The corresponding
expressions for quantum spins are similar but more
cumbersome. We do, however, include the results for quantum spins 
in the figures.

{\it Mechanisms.}---Our model Eq.\ (\ref{eq:Hamiltonian}) contains the
spin-orbit interaction only in the band structure. In this sense, the AHE in
this model is entirely {\it intrinsic}. Nevertheless, there are still several
mechanisms which contribute to the anomalous Hall conductivity, such as the
Berry-phase, the skew-scattering, and the side-jump contributions. 
Our diagrammatic results properly capture {\em all of these}
contributions. However, beyond low-order perturbation theory \cite{Sinitsyn}
it is difficult to disentangle the diagrammatic results. 
Specifically, the ladder diagrams in Fig.\ 1(a) contain 
Berry-phase and side-jump contributions. Third oder skew-scattering diagrams are
collected in Fig.\ 1(b). 
Also, in the present formalism all contributions
to the AHE come from the Fermi surface, while in  
Ref.~\cite{Jungwirth}  the Berry phase contribution has been attributed 
to the whole Fermi volume. 

\begin{figure}[t]
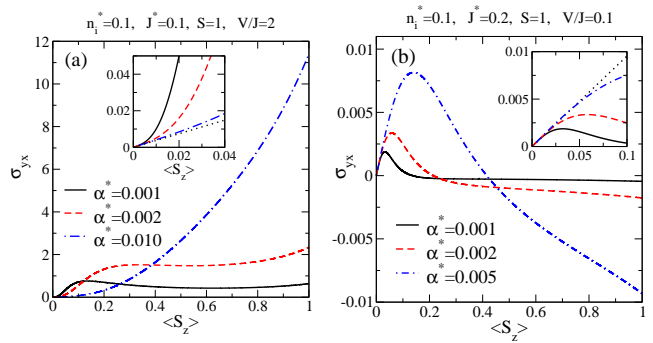

\begin{minipage}{.49\columnwidth}
\includegraphics[width=.95\columnwidth,clip=true]{V2S1new.eps}
\end{minipage}
\begin{minipage}{.49\columnwidth}
\includegraphics[width=.99\columnwidth,clip=true]{V2S1new2.eps}
\end{minipage}
\caption{(Color online.) Anomalous Hall conductivity for quantum spins with $S=1$ as
function of the polarization of the impurity spins 
$\langle S_z \rangle$. The insets display 
zoom-ins for small impurity magnetizations, where the
dotted lines indicate the $\alpha$-independent slope for small
polarization $\langle S_z \rangle$. All conductivities are plotted in
units of $e^2/(2 \pi)$ and we have used dimensionless units of the
form  
$n_i^*=n_i/k_F^2$,
$m^* = 1/2$,
$J^* = 2m J/\hbar^2$,
$V^* = 2m V/\hbar^2$,
$\alpha^* = 2m \alpha/(\hbar^2 k_F)$.} 
\label{fig:VJ}
\end{figure}

\begin{figure}[t]
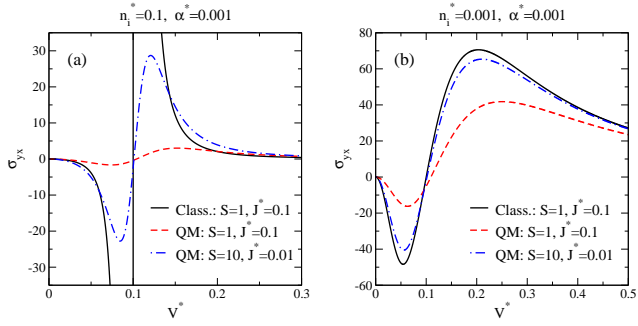

\begin{minipage}{.49\columnwidth}
\includegraphics[width=.95\columnwidth,clip=true]{bvc2a.eps}
\end{minipage}
\begin{minipage}{.49\columnwidth}
\includegraphics[width=.95\columnwidth,clip=true]{bvc3.eps}
\end{minipage}
\caption{(Color online.) Fully spin-polarized $\langle S_z \rangle =S$ anomalous Hall
  conductivity for classical $S=1$ spins (solid lines) and quantum
  spins (dashed and dashed-dotted lines) for constant $\alpha^*$, $n^*_i$ and $J^*$ as
  a function of $V^*$. Panel (a) shows the limit $\epsilon_J \gg \Delta_{\rm so}$
  and (b) the limit $\epsilon_{\rm so} \ll \epsilon_J \ll \Delta_{\rm so}$.
All conductivities are plotted in units of $e^2/(2 \pi)$ and
  we have used the dimensionless units of Fig.~\ref{fig:VJ}.} 
\label{fig:bvc}
\end{figure}

The skew-scattering diagrams collected in Fig.\ 1(b)
dominate in the limit of strong spin-orbit coupling
$\epsilon_{\rm so} \gg \epsilon_J, \epsilon_V$ where
\begin{eqnarray}
\sigma_{yx}^{I,s} \!\!  \approx  \!\! \sigma_{yx}^{I,l}  
\frac{\epsilon_{\rm so} (\epsilon_V^4-\epsilon_J^4)}
     {4\epsilon_J^2 \epsilon_V (\epsilon_V^2+3\epsilon_J^2)}.   
\end{eqnarray}
Here, we introduced the energy scale $\epsilon_V\equiv n_i V$, which measures the potential disorder strength.
By contrast, for weaker spin-orbit coupling $\epsilon_{\rm so} \ll \epsilon_J,\epsilon_V$, skew scattering and the ladder contributions can be of similar order of magnitude. For this reason, our plots always refer to the sum $\sigma_{yx}^{I,l}+\sigma_{yx}^{I,s}$, cf.\ Figs.~\ref{fig:VJ} and \ref{fig:bvc}. 

{\it Small impurity magnetization.}---In the Rashba model with pointlike disorder, the AHE vanishes in
the absence of magnetic scattering when both subbands are
occupied~\cite{Inoue,AHElong}. At small impurity magnetization
 we therefore
expect that the anomalous Hall conductivity is proportional to the
magnetization, i.e., to $h=n_i J \langle S_z \rangle$. However, in 
our model this holds only for very small polarization of the
impurity spins. This can be understood analytically from the small
magnetization expansion (for classical spins)  
\begin{eqnarray}
\sigma_{yx}^{I,l} \! & \approx &  \! \frac{e^2}{2\pi} 
\frac{\langle S_z \rangle}{S}
\frac{16 \epsilon_J^3}{3 \epsilon_V^2 + 7 \epsilon_J^2}
\left (\frac{\epsilon_V}{\epsilon_V^2 + \epsilon_J^2}
- \frac{2 \epsilon_J^2}{\epsilon_F ( 3\epsilon_V^2 + 7 \epsilon_J^2)} \right)\;, 
\nonumber \\
\sigma_{yx}^{I,s} \!& \approx & \! - \frac{e^2}{2 \pi} \frac{\langle S_z \rangle}{S}
\frac{18 \epsilon_{\rm so} \epsilon_J (\epsilon_J^2 -\epsilon_V^2)}
{(7 \epsilon_J^2 +  3 \epsilon_V^2)^2}\,.
\label{eq:slope}
\end{eqnarray}
Note that a nonzero exchange coupling $J$ is required for a non-vanishing AHE. 
Moreover, since the skew scattering contribution $\sigma_{yx}^{I,s}$
is negligible for $\epsilon_{\rm so} \ll \epsilon_J, \epsilon_V$, we obtain
the remarkable result  that in this limit, the slope of the anomalous Hall conductivity
becomes approximately {\em independent} of the
magnitude $\alpha$ of the spin-orbit interaction. 
Since a finite spin-orbit coupling is required for the existence of
the AHE in the first place, this implies that the regime over which the linear
magnetization dependence holds, must shrink with decreasing spin-orbit
coupling $\alpha$, entailing a strongly nonlinear behavior of the anomalous
Hall conductivity vs.\ polarization of the impurity spins $\langle S_z
\rangle$, as illustrated in Fig.~\ref{fig:VJ}.

Remarkably, we find that the AHE can even change sign as a function of
impurity magnetization (see Fig.~\ref{fig:VJ}(b)).
This behavior is reminiscent of sign changes found 
experimentally in other systems~\cite{Mihaly,Jungwirth}. 
Inspecting the expressions for the slope
[Eq.~(\ref{eq:slope})] and  for the conductivity at maximal 
spin polarization $\langle S_z \rangle=S$ [Eq.~(\ref{eq:classlimitS}) below], 
we find that for $\epsilon_{\rm so} \ll \epsilon_J, \epsilon_V$, 
where the skew scattering
contribution to the slope is negligible, the conductivity 
as a function of $\langle S_z \rangle$ changes sign 
for scalar impurity
potentials $V$ of intermediate strength, 
i.e., for $JS > V > JS \frac{\epsilon_J}{\epsilon_F} 
\frac{2 \epsilon_V^2 + 2 \epsilon_J^2}
{3 \epsilon_V^2 + 7 \epsilon_J^2}$.
Note also that the overall sign of the AHE depends on the sign of the
exchange coupling~$J$.

{\it Large impurity magnetization.}---The most intriguing behavior of
the anomalous Hall conductivity occurs in the limit, where the
impurity spins are near full polarization, i.e., for $\langle S_z
\rangle \simeq S$. We first focus on the regime of weak spin-orbit
coupling $\Delta_{\rm so} \ll h=n_i J \langle S_z \rangle$ (and thus
$\lambda_F = \sqrt{h^2 + \alpha^2 k_F^2} \approx h$). In this regime,
the scattering rate of the minority band 
\begin{equation}
\Gamma_+ \approx \Gamma - \Gamma_z \frac{h}{\lambda_F} = \frac{n_i}{2} \left( V^2 
- 2 V J \langle S_z \rangle \frac{h}{\lambda_F} + J^2 S (S\!\!+\!\!1) \right)
\end{equation}
{\it vanishes} at $V=JS$ for classical spins. This absence of scattering arises from {\em interference} between the potential and magnetic components of the impurity potential and implies a {\em divergence} of the anomalous Hall conductivity. Note that this absence of scattering holds for minority spins only. For the majority band, $\Gamma_-$ is {\em enhanced} due to the opposite sign of the interference term  $\propto \Gamma_z $ (see Eq.~(\ref{eq:Gammapm})).

Indeed, the anomalous Hall conductivity in 
the regime $\Gamma \ll \epsilon_J,\epsilon_V$   for classical fully
polarized magnetic impurities reads
\begin{equation}
\sigma_{yx}^{I,l} 
\approx \frac{e^2}{2 \pi}
\frac{4 \epsilon_{\rm so} \epsilon_F^2 \epsilon_V^2 (\epsilon_J^2+\epsilon_V^2)}
{\epsilon_J (\epsilon_V^2 - \epsilon_J^2)^3} \,,\,
\sigma_{yx}^{I,s}
\approx \sigma_{yx}^{I,l} \frac{2\epsilon_J^2}{\epsilon_J^2+\epsilon_V^2}  \,.
\label{eq:classlimitS}
\end{equation}
This expression obviously diverges for $\epsilon_V=\epsilon_J$ and also
implies a {\em sign change} of the anomalous Hall conductivity at
$\epsilon_V=\epsilon_J$. For quantum spins 
the true divergence is eliminated. Nevertheless, 
a strong enhancement of the AHE is predicted at full spin
polarization for $\epsilon_V\approx \epsilon_J$. 
As expected, the result for quantum spins 
approaches the classical divergence as the spin increases 
(see Fig.~\ref{fig:bvc}).  

In addition, the  divergence in the approximate expression,  
Eq.~\eqref{eq:classlimitS}, is also cut off by a finite spin-orbit coupling
$\alpha$. However, the AHE remains strongly enhanced near
full polarization $\langle S_z \rangle =S$ even when $\epsilon_{\rm so} \ll
\epsilon_J \ll \Delta_{\rm so}$. In this regime of strong spin-orbit
scattering, we obtain  
\begin{eqnarray}
\sigma_{yx}^{I,l} \!\! \approx \!\! \frac{e^2}{2 \pi}
\frac{4 \epsilon_J^3 \epsilon_V^2 (\epsilon_V^2 \!-\!\epsilon_J^2)}
{\epsilon_{\rm so} (\epsilon_V^2\!+\!\epsilon_J^2)^2(\epsilon_V^2\!+\!3\epsilon_J^2)} \,,\,
\sigma_{yx}^{I,s} \!\! \approx \!\!
 \sigma_{yx}^{I,l} \frac{2\epsilon_J^2}{\epsilon_V^2 \!+\! 3\epsilon_J^2}  
\label{eq:classlimitSsmallerh}
\end{eqnarray}
for classical spins. Obviously, the true divergence of the anomalous
Hall conductivity at $\epsilon_V=\epsilon_J$ has disappeared 
but a pronounced maximum of the anomalous Hall conductivity survives at
$\epsilon_V \approx 2.2 \epsilon_J$ as can be
seen in Fig.~\ref{fig:bvc}(b) (black line). Again quantum spins behave very
similarly, and the classical result is recovered for large spins.
Interestingly, for fully polarized quantum spins, $\langle S_z \rangle =S$,
the anomalous Hall conductivity becomes maximal in the crossover
regime between the two limits discussed above, i.e., for
$\epsilon_J \approx \Delta_{\rm so}$.

{\it Conclusions.}---We have investigated the effects of magnetic
impurities on the anomalous Hall effect, and uncovered rich and
unexpected behavior of the anomalous Hall conductivity. 
We find a highly nonlinear dependence on the spin magnetization, 
sign changes of the AHE as function of magnetization, as well as a resonant
enhancement due to interference between potential and magnetic
scattering from the magnetic impurities. 

Although our work is motivated by the ubiquitous presence of magnetic
impurities in spintronics materials, we have focused here on a thorough
theoretical analysis of a simple model system, namely a two-dimensional electron
system with Rashba spin-orbit coupling.  
A more realistic description of the band structure of real spintronics
materials remains an important task for future research. Indeed, recent
experimental work~\cite{Mihaly} on the diluted magnetic semiconductor
(In,Mn)Sb exhibits an intriguing sign change of the anomalous Hall
conductivity as function of the impurity magnetization which is interpreted in
terms of the Berry-phase contribution to the AHE~\cite{Jungwirth}. Our work
shows that the occurrence of such sign changes is much more generic and a
conclusive interpretation of the experimental data must await more detailed
research. 

{\it Acknowledgments.}---This work was supported by a M\"OB-DAAD grant as well
as SPP 1285 of the DFG. G.Z. Has been supported by the Hungarian Grants
OTKA T046303 and NF061726.

\end{document}